\pdfoutput=1
\documentclass[11pt]{article}

\usepackage[final]{acl}

\usepackage{times}
\usepackage{latexsym}
\usepackage{tablefootnote}
\usepackage{graphicx}
\usepackage{multirow}

\usepackage[T1]{fontenc}
\usepackage[utf8]{inputenc}

\usepackage{microtype}

\usepackage{inconsolata}

\title{LLM-Augmented Retrieval: Enhancing Retrieval Models Through Language Models and Doc-Level Embedding}

\author{Mingrui Wu \\
  Meta \\
  \texttt{mingruiwu@meta.com} \\\And
  Sheng Cao \\
  Meta \\
  \texttt{rickcao@meta.com} \\}

\begin{document}
\maketitle
\begin{abstract}
Recently embedding-based retrieval or dense retrieval have shown state of the art results, compared with traditional sparse or bag-of-words based approaches. This paper introduces a model-agnostic doc-level embedding framework through large language model (LLM) augmentation. In addition, it also improves some important components in the retrieval model training process, such as negative sampling, loss function, etc. By implementing this LLM-augmented retrieval framework, we have been able to significantly improve the effectiveness of widely-used retriever models such as Bi-encoders (Contriever, DRAGON) and late-interaction models (ColBERTv2), thereby achieving state-of-the-art results on LoTTE datasets and BEIR datasets.
\end{abstract}

\section{Introduction}
The Bi-encoder \citep{karpukhin2020dense} is a type of neural network architecture that is widely used in information retrieval. It consists of two encoders, typically in the form of transformer models \citep{vaswani2017attention}, which encode an vector representation for user queries and potential documents or passages respectively. These two encoders can be shared or using two separate models. The similarity between these two embedding vectors can then be computed, often using dot product or cosine similarity, to determine the relevance of the document or passage to the user's query.

\begin{figure}[h!]
    \centering
    \includegraphics[width=1\linewidth]{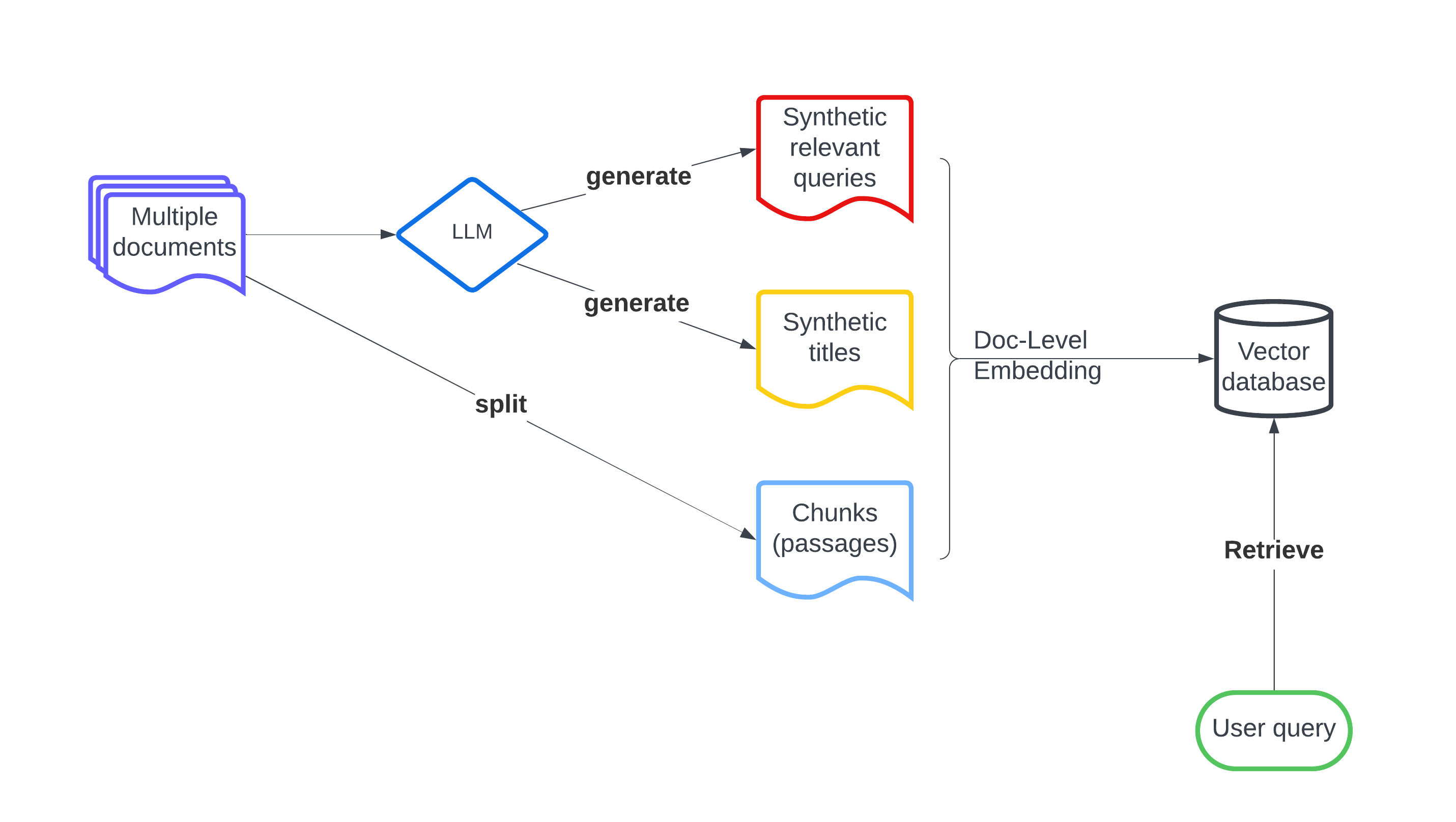}
    \caption{Overall view on LLM-augmented retrieval framework. Synthetic relevant queries and synthetic titles are generated from LLM and then assembled into doc-level embedding together with chunks (passages) split from the original document. The final retrieval is based on the similarity between user query and the doc-level embedding.}
    \label{fig:relevance_framework}
\end{figure}

Cross-encoders \citep{nogueira2019passage}, unlike bi-encoders, amalgamate the inputs at an early stage, allowing for a more intricate interaction between user queries and documents. Here the user query and the document are concatenated, based on which a joint embedding vector is computed. The joint embedding vector is then used to make predictions, such as the relevance of a document to a query in an information retrieval task. Cross-encoders often outperform bi-encoders in tasks requiring a nuanced understanding of the interplay between inputs.

Late-interaction models, such as ColBERT \citep{khattab2020colbert}, ColBERTv2 \citep{santhanam2021colbertv2} or SPALDE++ \citep{formal2022distillation}, are model architectures that hybrids cross-encoder models and bi-encoder models. Queries and documents are independently encoded into token-level vector representations. So in some sense, this is a bag of embedding vectors model. The interaction between these representations, which constitutes the “late interaction”, involves computing the cosine similarity or dot product scores over the token-level vector embedding. 

All the model architectures require informative embedding of user queries and target documents. While we cannot control the user queries during retrieval tasks, we hypothesize that enriching the embedding of documents can improve the quality and robustness of text retrieval.
				
This work makes the following contributions:
\begin{enumerate}
\item We propose LLM-augmented retrieval, a model-agnostic framework that enriches the contextual information in the vector embedding of documents to improve the quality and robustness of existing retrievers.
\item We propose doc-level embedding, which combines more contextual information in the context embedding.
\item We evaluate this framework across different models and a wide range of datasets, establishing state-of-art quality beyond original models.
\item We propose improvements to some key components in retrieval model training, such as negative sampling, loss function, etc. 
\end{enumerate}

\section{Background \& Related Work}
To the best of our knowledge, there’s no existing work on constructing a unified LLM-augmented framework for doc-level embedding. Therefore, we summarized some related knowledge that’s important to our work.

\subsection{Data Augmentation in Information Retrieval}
Data augmentation is a widely used technique in information retrieval training. In Contrastive Learning \citep{izacard2021unsupervised}, inverse cloze task, independent cropping, as well as random word deletion, replacement or masking are introduced to enrich the diversity of training data. While in training the DRAGON model, \citep{lin2023train}, not only query augmentation by query generation models are studied, but also label augmentation methods with diverse supervision are discussed and compared.

Large pre-trained language models are good at generating text data of high quality \citep{anaby2020not, papanikolaou2020dare, yang2020generative, kumar2020data, schick2021generating, meng2022generating}. Some past work has been done to utilize the generation capability of language models to create synthetic training data for retriever models \citep{bonifacio2022inpars, wang2023improving}.

\subsection{Pseudo Queries Generation}
Pre-generated pseudo queries are proved to be effective in improving retrieval performance. Previously the similarity of pseudo-queries and user-queries are calculated through BM25 or BERT model in determining the final relevance score of the query to document through relevance score fusion \citep{chen2021contextualized, wen2023offline}. An alternative method to generate pseudo queries is to generate pseudo query embedding through K-means clustering algorithm \citep{tang2021improving}.

\subsection{Retrieval Augmented Generation (RAG)}
In a Retrieval Augmented Generation (RAG) system \citep{lewis2020retrieval}, a dedicated retriever module is introduced to retrieve relevant documents from a set of corpus based on the input query, and the retrieved documents are then integrated by a language model as part of the context, in order to refine its final response generation. This approach can be applied to not only decoder-only models but also encoder-decoder architectures \citep{yu2022retrieval}.

\subsection{Similarity Scores}
Having the embedding vectors of input query and target document, we need to compute the similarity score between these two vectors to determine the relevance between them \citep{jones1987pictures}. Dense retrievers like Contriever \citep{izacard2021unsupervised} and DRAGON \citep{lin2023train} use dot products to measure the similarity of query and document in the text embedding space. SPLADE \citep{formal2021splade} and SPLADEv2 \citep{formal2021spladev2} use dot products as well while ColBERT \citep{khattab2020colbert} and ColBERTv2 \citep{santhanam2021colbertv2} use cosine similarity scores.
The main difference between cosine similarity and dot product is the sensitivity to the magnitude of the embedding vectors.  And dot product is computationally cheaper and more stable. 

\subsection{Training}
The contrastive InfoNCE loss \citep{izacard2021unsupervised, lin2023train} is widely adopted in dense retrieval training and is defined as:
\begin{equation}
\resizebox{.85\hsize}{!}{$L(q, d_{+}) = - \frac{exp(s(q, d_{+})/\tau)}{exp(s(q, d_{+})/\tau) + \sum_{i=1}^{K}exp(s(q, d_{i})/\tau)}$}
\end{equation}
\noindent
where $\tau$ is the temperature, $s$ is the similarity score function, $q$ is the input query, $d_i$ is any candidate document and $d_+$ is the relevant document. Other popular loss functions include point-wise Binary Cross Entropy \citep{sun2023chatgpt}, list-wise Cross Entropy \citep{bruch2021alternative}, RankNet \citep{10.1145/1102351.1102363} and LambdaLoss \citep{wang2018lambdaloss}. Pointwise Binary Cross Entropy is calculated based on each of the query-document pairs independently. List-wise Cross-Entropy loss is widely used in passage ranking and minimizes list-wise softmax cross entropy on all passages. RankNet is used to calculate the pair-wise relevant order of passages. 

\section{LLM-augmented Retrieval}
In this section, we first discuss the components of the LLM-augmented retrieval framework. After that, we explain how this framework can be adapted to different retriever model architectures. In particular, we propose doc-level embedding for bi-encoders and late-interaction encoders under the LLM-augmented retrieval framework and demonstrate how it is applied to enhance the end-to-end retrieval quality. 

\subsection{LLM-Augmented Retrieval Framework}

\subsubsection{Synthetic Relevant Queries}
The inspiration for this concept was drawn from web search techniques \citep{xue2004optimizing, guo2009click, guo2009efficient, chuklin2022click}. To illustrate the idea, let us consider an example on a user query of "MIT". Without prior knowledge, it is challenging to figure out that "Massachusetts Institute of Technology" and "MIT" are equivalent. Nevertheless, in web search, we can observe that the home page of "Massachusetts Institute of Technology" has received numerous clicks from the query "MIT", allowing us to infer that the home page of "Massachusetts Institute of Technology" must be closely associated with the query "MIT". On the other side, we typically don’t have click data for each user query in the scenario of contextual retrieval. However, large language models are good at generating synthetic queries \citep{anaby2020not, papanikolaou2020dare, yang2020generative, kumar2020data, schick2021generating, meng2022generating} so we can use the synthetic queries as proxy “click data” to direct user queries to related documents.

An important point is that in traditional retrieval tasks, we are using similarity to express relevance \citep{jones1987pictures}. The similarity score is mathematically defined as either dot product or cosine of the encoded vectors of user query and documents. However, sometimes this similarity score might not reflect semantic relevance \citep{rao2019bridging}. For example, “Who is the first president of the United States?” might be very close, in terms of similarity score, to “Who became the first president of America?”. But our target answer might be a Wiki page or autobiography about “George Washington”, whose similarity score to the query is not that high. While if we use Washington’s autobiography to create synthetic queries, “Who became the first president of America?” may be one of them. The user query “Who is the first president of the United States?” can easily match to the relevant query through similarity scores. And the latter points to the target document (Washington’s autobiography). Therefore the generated synthetic queries express the semantic of the original document from different angles which are helpful to match the relevant queries. 

% \begin{figure}[h!]
%     \centering
%     \includegraphics[width=1\linewidth]{relevance_1.png}
%     \caption{In traditional retrieval tasks, we are using similarity to express relevance.}
%     \label{fig:enter-label}
% \end{figure}

\begin{figure}[h!]
    \centering
    \includegraphics[width=1\linewidth]{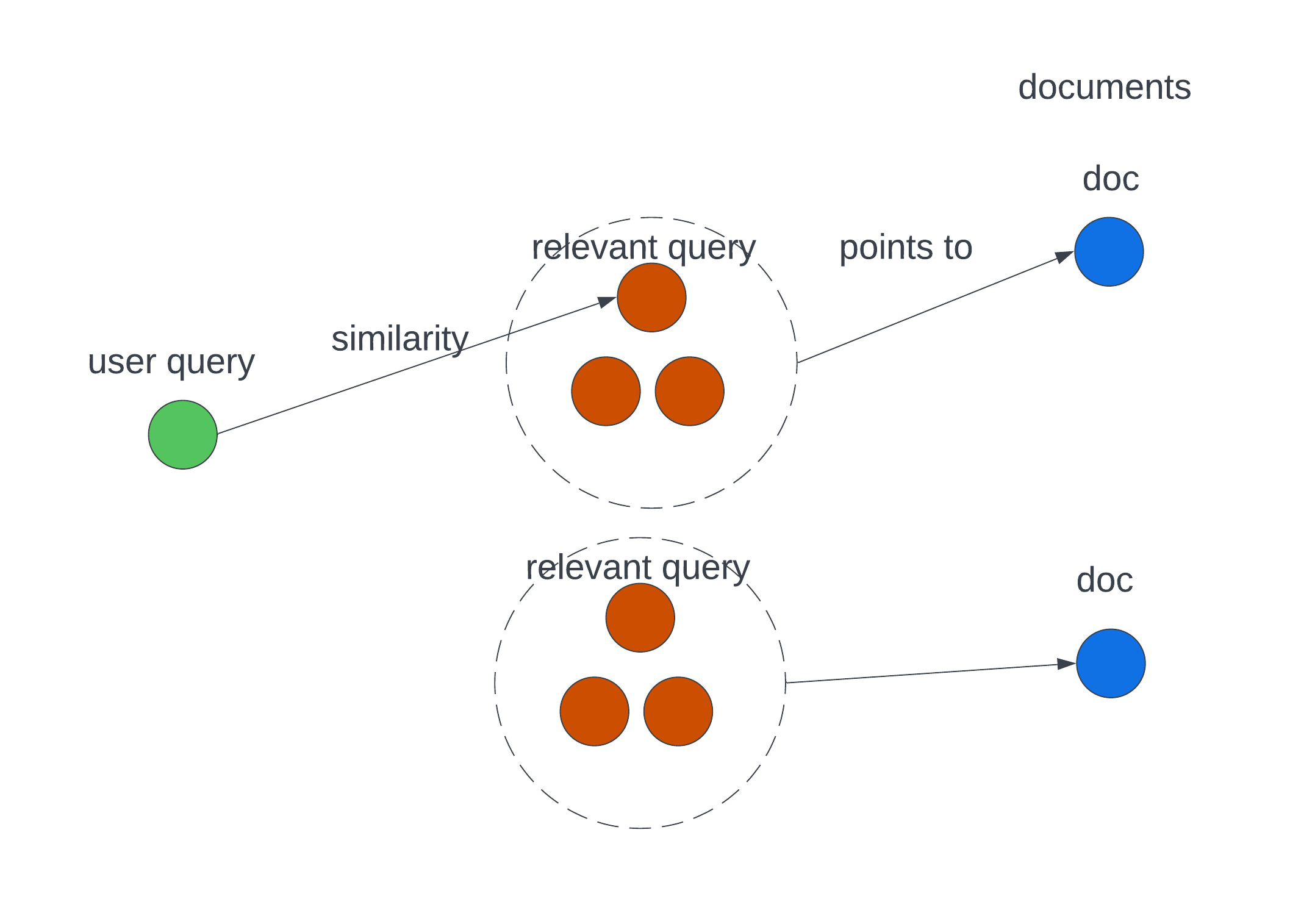}
    \caption{Through synthetic relevant queries, the relevance relationship is not solely expressed by the similarity now but also expressed by the augmentation steps of the large language models}
    \label{fig:relevance_full}
\end{figure}

\subsubsection{Title}
The title of a document plays a crucial role in determining its relevance and usefulness to a user's query. The title is often the first piece of information that a user sees when searching for documents, and it can greatly influence their decision to click on a particular result. A well-crafted title can provide important context and keywords that help users quickly understand the content and purpose of a document. If the original document has a title, we can use them directly. If it does not, we can leverage the large language models to generate a synthetic title for a that document.

\subsubsection{Chunks (Passages)}
Chunking refers to the process of dividing a large document or piece of text into smaller, more manageable units. These units, known as "chunks," or "passages", are typically created by grouping together related pieces of information. Due to the limitation of the context window of retriever models, (in other words, max length of the model input), we typically divide a long document into several chunks whose number of tokens is below the context window limit. The chunk data is from original documents and is not from LLM-augmentation. The optimal chunking size is different for various retriever models. For bi-encoders like Contriever and DRAGON, we found the optimal chunking size to be 64 after empirical studies. For token-level late-interaction models such as ColBERT and ColBERTv2, since it’s already calculating the similarity score at token-level, it’s not necessary to chunk the original documents unless the context window limit is reached.

\subsection{Doc-level embedding}
In this section, we will first introduce the high level idea of doc-level embedding for information retrieval, and then use bi-encoders and token-level late-interaction models to illustrate how the doc-level embedding can be adaptive to different retriever model structures.

\textbf{Document fields.} For convenience, we call the above mentioned information source, synthetic queries, title and chunk, \textit{fields} of a document. These fields express the semantic of the original document from different angles and will be composed into the doc-level embedding of a document which is static and can be pre-computed and cached for information retrieval. Embedding indexes can be pre-built to speed up the retrieval inference, and each doc-level embedding points to the original document.

\begin{figure}[h!]
    \centering
    \includegraphics[width=1\linewidth]{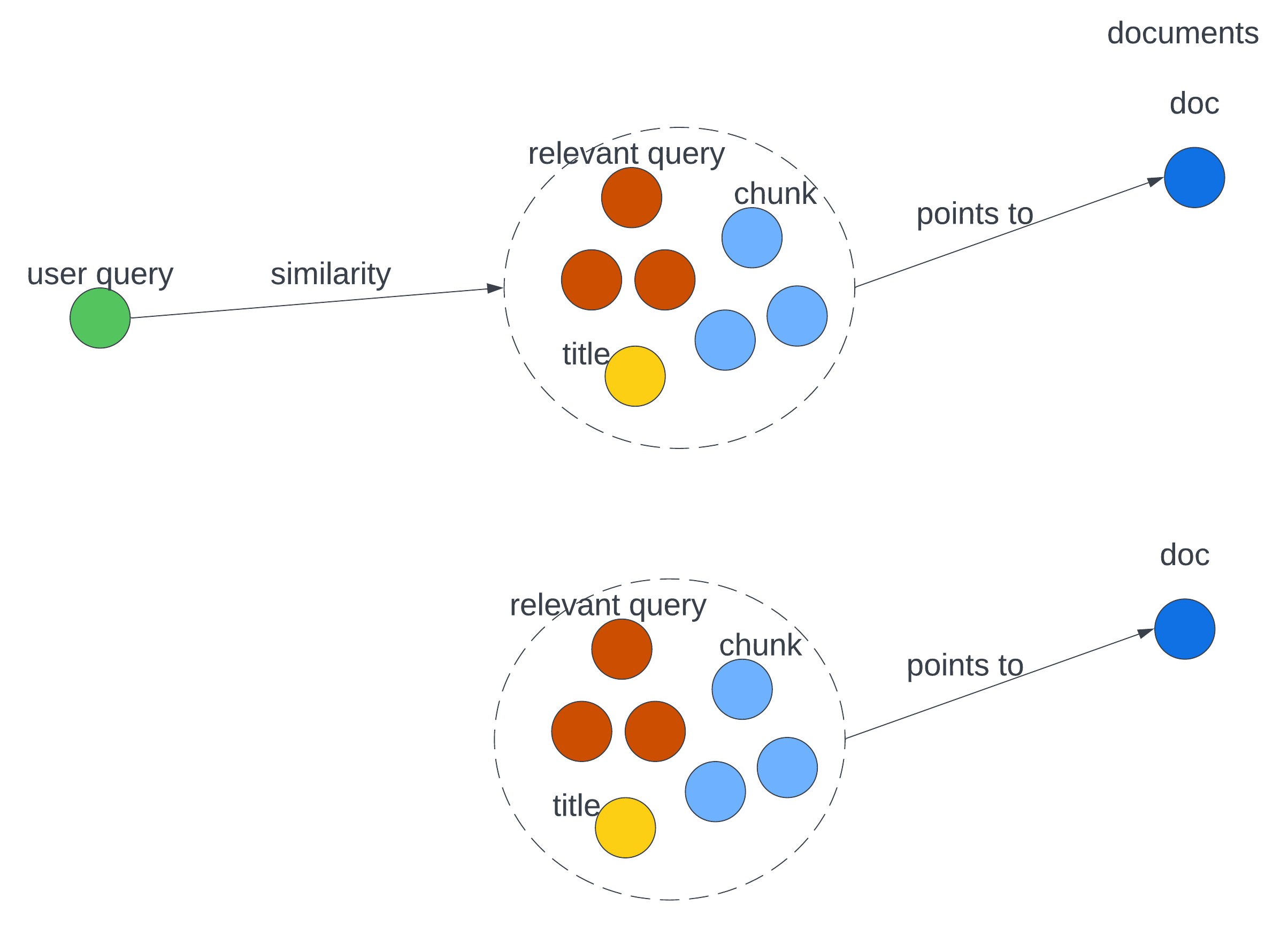}
    \caption{The graphic representation of "relevance" in doc-level embedding}
    \label{fig:enter-label}
\end{figure}

\subsubsection{For Bi-encoders}
Bi-encoders are typically “Two-Tower” model structures. Given a query and a document, a query encoder and a doc encoder are applied to compute the embedding vectors for the query and the document respectively. Then these two embedding vectors are fed into the dot products (or cosine similarity) to compute the similarity scores between the query and the document. As we want to enrich the document embedding vectors by injecting the synthetic queries and titles, we propose to compute the similarity as the following: 
\begin{equation}
\resizebox{.85\hsize}{!}{$sim(q, d) = max_{i}s(q, c_{i}) + \sum_{f}w_f \times s(q, e_{f})$}
\end{equation}
\noindent
The first term on the right hand side calculates the maximum similarity score of query chunk embedding pairs, where $s$ is the similarity score function, $q$ is the embedding vector of the input query, $c_i$ is the embedding vector of the $i$-th chunk in the document. This term is commonly used in current embedding based retrievals, which determines the similarity between a query and a document based on the query and the most relevant chunk in the document.

The second term is innovative and considers more information, where $e_f$ is the embedding vector of each document field. The similarity scores between the query embedding and each field embedding are computed and combined together, each with a field weight parameter $w_f$. As described before, those document fields include the synthetic query, title and chunk fields. 

Now let's consider how to compute the query field embedding $e_f$ for each field. For the title field, which just contains one phase or sentence, it is straightforward. We just apply the doc encoder to compute the embedding vector of the title as the title field embedding. 

For the chunk field which may contain multiple doc chunks, we can compute the embedding vector of each chunk with the doc encoder. The question is how to combine these embedding vectors to represent the whole document? Actually people have already considered a similar problem for sentence embedding: having the embedding vectors of all the tokens in one sentence, how to come up with a representation of the whole sentence? A simple but effective idea was proposed in \citep{Sanjeev2017} for this problem. Here we adapt this idea to the doc embedding problem. Namely we compute the average of all the chunk embedding vectors as the chunk field embedding. Similarly, for the synthetic query field, we compute the embedding vector of each query with the query encoder, and then compute the average of these embedding vectors as the query field embedding. This simple approach works very well in our experiments, while clearly more advanced approaches can be explored here in the future.

Furthermore, as the similarity function is linear\footnote{Dot product is linear. Cosine similarity is also linear if we normalize the embedding vectors to unit length.}, the above equation can be simplified as the following:
\begin{equation}
\resizebox{.85\hsize}{!}{$sim(q, d) = max_{i}s(q, c_{i}+ \sum_{f}w_f \times e_{f})$}
\end{equation}

Therefore, we can treat $c_i+\sum_f w_f \times e_f$ as the chunk embedding vector of each chunk $c_i$ of the original document, and still apply the algorithms like approximate nearest neighbors \citep{indyk1998approximate} to retrieve the most relevant documents. 

\subsubsection{For Token-Level Late-Interaction Models}
Instead of using a single embedding vector for query and a single embedding vector for each document, late-interaction models such as ColBERT and ColBERTv2 use token-level embedding and the embedding vectors of all the tokens are kept and will participate in computing the similarity score between the query and document.
\begin{equation}
{sim(q, d) = \sum_{i}max_{j}s(q_i, t_{j})}
\end{equation}
\noindent
where $q_i$ and $t_j$ are token-level embedding vectors for the input query and document respectively. Therefore, for each query token, the most similar token from the document is identified, and their similarity score is recorded. All these scores are summed up over all the query tokens to obtain the overall similarity between the query and the document. Since the similarity score calculation is done at token-level, we can concatenate the synthetic queries and titles to the original document passages. After that, we decide whether to chunk the concatenated documents if the number of tokens reaches the context window limit.

\section{Experiments}
In this section we will introduce our experiment set up, the datasets and the models used in our experiments. 

\subsection{Datasets}
\textbf{BEIR Data}  The BEIR (Benchmark for Evaluating Information Retrieval) dataset \citep{thakur2021beir}, provides a comprehensive benchmark for evaluating and comparing different IR models, particularly in the context of out-of-domain tests. BEIR is designed to address the limitations of other previous datasets by offering a more diverse and extensive range of queries and passages. It encompasses a wide array of topics, thereby facilitating a more robust and comprehensive evaluation of IR models.\\

\noindent
\textbf{LoTTE Data} LoTTE data \citep{santhanam2021colbertv2} is a dataset specifically designed for Long-Tail Topic-stratified Evaluation. The primary focus of LoTTE is on natural user queries that are associated with long-tail topics, which may not be adequately covered by an entity-centric knowledge base such as Wikipedia. LoTTE is composed of 10 distinct test sets, each containing 500 -- 2,000 queries and 100,000 -- 2,000,000 passages. These test sets are divided by topic, and each is accompanied by a validation set comprising related yet disjoint queries and passages. In this experiment we are using the test split sorely for evaluation.

\subsection{Models}
\textbf{Contriever} For Contriever we use the Roberta-base \citep{liu2019roberta} model architecture, the checkpoint is trained on Wiki passages \citep{karpukhin2020dense} and CC100 \citep{conneau2019unsupervised} data through Contrastive Learning. It has 125M parameters, a context window of 512 tokens, with 12 layers, 768 hidden dimensions, and 12 attention heads. We use the single Roberta-base model for query encoder and context encoder, namely the “Two Tower” of the bi-encoder is shared in model architecture. \\

\noindent
\textbf{DRAGON} For DRAGON we also use the Roberta-base model architecture, the checkpoint is trained by the author and released publicly. Different from the Contriever model, DRAGON has separate Roberta-base models for query encoder and context encoder.\\

\noindent
\textbf{ColBERTv2}  For ColBERTv2, we use bert-base-uncased model architecture, the same as the default settings in the original paper. It has 110M parameters and a context window of 256 tokens, with 12 layers, 768 hidden dimensions, and 12 heads. The checkpoint is trained on the MSMARCO dataset \citep{nguyen2016ms} from the author’s checkpoint.

\subsection{Implementation Details}
We choose open source Llama-70B \citep{touvron2023llama, touvron2023llama2} for synthetic query generation and title generation. The prompt templates used for generating synthetic queries and titles are in Table \ref{tab:prompt_query} and \ref{tab:prompt_title}.

For Bi-encoders, we implemented the doc-level embedding as above mentioned and chose $w_{query}$=1.0, $w_{title}$=0.5, $w_{chunk}$=0.1 for the Contriever model and $w_{query}$=0.6, $w_{title}$=0.3, $w_{chunk}$=0.3 for the DRAGON model. We chose $chunk\_size$=64 after empirical experiments and found 64 normally performed the best in retrieval results for Bi-encoders. Note these hyper-parameters of field weights are not fully optimized. They are chosen according to the performance of Bi-encoder models on the single LoTTE-lifestyle-forum dataset, and then fixed for the evaluations on all the other datasets. 

For ColBERTv2, as mentioned previously, we concatenate the title with all the synthetic queries for each document and make it an additional “passage” of the original document. Thus there’s no field weights hyper-parameters in these experiments. There could be other better assembling methods for composing the doc-level embedding under a late-interaction model architecture. We set index\_bits=8 when building the ColBERT index. 

\section{Results}
The result on LoTTE and BEIR data for all three models can be seen in Table \ref{tab:contriever_results}, \ref{tab:dragon_results} and \ref{tab:colbertv2_results}. We can observe that the LLM augmented retrieval and doc-level embeddings boost the recall@3 and recall@10 of Bi-encoders (Contriever and DRAGON) significantly. For token-level late-interaction models (ColBERTv2), the increase on the LoTTE and BEIR Dataset is still clear, although not as much as on Bi-encoders. We hypothesis this due to that the baselines for token-level late-interaction models are much higher than those of Bi-encoders. 

Moreover, the performance of the LLM-augmented Contriver has exceeded the performance of the vanilla DRAGON in most datasets. Similarly, the LLM-augmented DRAGON even exceeds vanilla ColBERTv2 on BEIR-ArguAna, BEIR-SciDocs and BEIR-CQADupstack-English datasets and greatly reduce the performance gap in the remaining datasets, although ColBERTv2 introduces a more complex late-interaction architecture than DRAGON. Therefore, we can see that, after enriching the embedding of documents with LLM augmentation, we can greatly improve the retriever models recall performance without further fine-tuning.

\begin{table*}[ht!]
\centering
\resizebox{\textwidth}{!}{%
\begin{tabular}{ p{0.15\linewidth} p{0.1\linewidth} p{0.1\linewidth} p{0.1\linewidth} p{0.1\linewidth} p{0.1\linewidth} p{0.1\linewidth} p{0.1\linewidth} p{0.1\linewidth} p{0.1\linewidth} p{0.1\linewidth} p{0.1\linewidth}}
\hline
\multicolumn{12}{c}{\textbf{LoTTE Dataset}} \\
\hline
Model & Recall & Lifestyle Forum & Lifestyle Search & Recreation Forum & Recreation Search & Science Forum & Science Search & Technology Forum & Technology Search & Writing Forum & Writing Search\\
\hline
\multirow{2}{*}{Contriever} & recall@3 & 0.4366 & 0.3358 & 0.3486 & 0.1948 & 0.1046 & 0.1005 & 0.1826 & 0.1242 & 0.3950 & 0.2745 \\
& recall@10 & 0.6149 & 0.4690 & 0.4895 & 0.2857 & 0.1706 & 0.1637 & 0.3174 & 0.1896 & 0.5390 & 0.3950 \\
\hline
\multirow{2}{*}{Contriever*} & recall@3 & \textbf{0.6244} & \textbf{0.6021} & \textbf{0.5455} & \textbf{0.4610} & \textbf{0.2395} & \textbf{0.2901} & \textbf{0.3663} & \textbf{0.3557} & \textbf{0.5970} & \textbf{0.5724} \\
& recall@10 & \textbf{0.7622} & \textbf{0.7821} & \textbf{0.6948} & \textbf{0.6320} & \textbf{0.3570} & \textbf{0.4684} & \textbf{0.5494} & \textbf{0.5017} & \textbf{0.7365} & \textbf{0.6919} \\
\hline
\end{tabular}
}
\resizebox{\textwidth}{!}{%
\begin{tabular}{ p{0.15\linewidth} p{0.1\linewidth} p{0.1\linewidth} p{0.1\linewidth} p{0.1\linewidth} p{0.1\linewidth} p{0.1\linewidth} p{0.12\linewidth} p{0.12\linewidth}}
\hline
\multicolumn{9}{c}{\textbf{BEIR Dataset}} \\
\hline
Model & Recall & ArguAna & FIQA & Quora & SciDocs & SciFact & CQADupstack English & CQADupstack Physics \\
\hline

\multirow{2}{*}{Contriever} & recall@3 & \textbf{0.2589} & 0.1895 & 0.8654 & 0.1580 & 0.5410 & 0.2261 & 0.1723 \\
& recall@10 & 0.5206 & 0.2993 & 0.9463 & 0.2950 & 0.6934 & 0.3089 & 0.2551 \\
\hline
\multirow{2}{*}{Contriever*} & recall@3 & 0.2468 & \textbf{0.3690} & \textbf{0.8687} & \textbf{0.2440} & \textbf{0.5996} & \textbf{0.3822} & \textbf{0.3417} \\
& recall@10 & \textbf{0.5825} & \textbf{0.5174} & \textbf{0.9517} & \textbf{0.4030} & \textbf{0.7259} & \textbf{0.5025} & \textbf{0.4658} \\
\hline
\hline
\end{tabular}
}
\caption{Results on Contriever: The performance of LLM-augmented Contriver has greatly exceeded the vanilla Contriever on both LoTTE and BEIR dataset, and even exceeds the performance of the vanilla DRAGON in most datasets. * means base model plus the doc-level embedding ($w_{query}$=1.0, $w_{title}$=0.5, $w_{chunk}$=0.1). }
\label{tab:contriever_results}
\end{table*}

\begin{table*}[ht!]
\centering
\resizebox{\textwidth}{!}{%
\begin{tabular}{ p{0.15\linewidth} p{0.1\linewidth} p{0.1\linewidth} p{0.1\linewidth} p{0.1\linewidth} p{0.1\linewidth} p{0.1\linewidth} p{0.1\linewidth} p{0.1\linewidth} p{0.1\linewidth} p{0.1\linewidth} p{0.1\linewidth}}
\hline
\multicolumn{12}{c}{\textbf{LoTTE Dataset}} \\
\hline
Model & Recall & Lifestyle Forum & Lifestyle Search & Recreation Forum & Recreation Search & Science Forum & Science Search & Technology Forum & Technology Search & Writing Forum & Writing Search\\
\hline
\multirow{2}{*}{DRAGON} & recall@3 & 0.5270 & 0.5598 & 0.4560 & 0.4253 & 0.2578 & 0.2601 & 0.2854 & 0.3591 & 0.5300 & 0.5798 \\
& recall@10 & 0.6798 & 0.7035 & 0.5949 & 0.5325 & 0.3704 & 0.3938 & 0.4232 & 0.5101 & 0.6675 & 0.7311 \\
\hline
\multirow{2}{*}{DRAGON*} & recall@3 & \textbf{0.6883} & \textbf{0.7625} & \textbf{0.6079} & \textbf{0.6472} & \textbf{0.3099} & \textbf{0.4498} & \textbf{0.4192} & \textbf{0.5285} & \textbf{0.6520} & \textbf{0.7031} \\
& recall@10 & \textbf{0.8172} & \textbf{0.8911} & \textbf{0.7468} & \textbf{0.7944} & \textbf{0.4427} & \textbf{0.6062} & \textbf{0.6038} & \textbf{0.7097} & \textbf{0.7725} & \textbf{0.8170} \\
\hline
\end{tabular}
}
\resizebox{\textwidth}{!}{%
\begin{tabular}{ p{0.15\linewidth} p{0.1\linewidth} p{0.1\linewidth} p{0.1\linewidth} p{0.1\linewidth} p{0.1\linewidth} p{0.1\linewidth} p{0.12\linewidth} p{0.12\linewidth}}
\hline
\multicolumn{9}{c}{\textbf{BEIR Dataset}} \\
\hline
Model & Recall & ArguAna & FIQA & Quora & SciDocs & SciFact & CQADupstack English & CQADupstack Physics \\
\hline

\multirow{2}{*}{DRAGON} & recall@3 & 0.1408 & 0.3327 & 0.8465 & 0.1800 & 0.4743 & 0.2605 & 0.1877 \\
& recall@10 & 0.4040 & 0.4514 & 0.9419 & 0.3260 & 0.5996 & 0.3599 & 0.2916 \\
\hline
\multirow{2}{*}{DRAGON*} & recall@3 & \textbf{0.3663} & \textbf{0.4255} & \textbf{0.8638} & \textbf{0.3040} & \textbf{0.6610} & \textbf{0.4618} & \textbf{0.3936} \\
& recall@10 & \textbf{0.6764} & \textbf{0.5635} & \textbf{0.9527} & \textbf{0.4800} & \textbf{0.7710} & \textbf{0.5662} & \textbf{0.5342} \\
\hline
\hline
\end{tabular}
}
\caption{Results on DRAGON: The performance of LLM-augmented DRAGON has greatly exceeded the vanilla DRAGON on both LoTTE and BEIR dataset, and even exceeds vanilla ColBERTv2 on BEIR-ArguAna, BEIR-SciDocs and BEIR-CQADupstack-English datasets, as well as greatly reduces the performance gap in the remaining datasets. * means base model plus the doc-level embedding ($w_{query}$=0.6, $w_{title}$=0.3, $w_{chunk}$=0.3). }
\label{tab:dragon_results}
\end{table*}

\begin{table*}[ht!]
\centering
\resizebox{\textwidth}{!}{%
\begin{tabular}{ p{0.15\linewidth} p{0.1\linewidth} p{0.1\linewidth} p{0.1\linewidth} p{0.1\linewidth} p{0.1\linewidth} p{0.1\linewidth} p{0.1\linewidth} p{0.1\linewidth} p{0.1\linewidth} p{0.1\linewidth} p{0.1\linewidth}}
\hline
\multicolumn{12}{c}{\textbf{LoTTE Dataset}} \\
\hline
Model & Recall & Lifestyle Forum & Lifestyle Search & Recreation Forum & Recreation Search & Science Forum & Science Search & Technology Forum & Technology Search & Writing Forum & Writing Search\\
\hline
\multirow{2}{*}{ColBERTv2} & recall@3 & 0.6988 & 0.7927 & 0.6344 & 0.6677 & 0.3932 & 0.5073 & 0.4496 & 0.5940 & 0.6960 & 0.7423 \\
& recall@10 & 0.8087 & 0.8911 & 0.7498 & 0.7868 & 0.5285 & 0.6613 & 0.6292 & 0.7315 & 0.8050 & 0.8366 \\
\hline
\multirow{2}{*}{ColBERTv2*} & recall@3 & \textbf{0.7308} & \textbf{0.8003} & \textbf{0.6753} & \textbf{0.7100} & \textbf{0.4026} & \textbf{0.5024} & \textbf{0.4626} & \textbf{0.5956} & \textbf{0.7145} & \textbf{0.7544} \\
& recall@10 & \textbf{0.8447} & \textbf{0.9107} & \textbf{0.7862} & \textbf{0.8268} & \textbf{0.5558} & \textbf{0.6726} & \textbf{0.6517} & \textbf{0.7383} & \textbf{0.8260} & \textbf{0.8571} \\
\hline
\end{tabular}
}
\resizebox{\textwidth}{!}{%
\begin{tabular}{ p{0.15\linewidth} p{0.1\linewidth} p{0.1\linewidth} p{0.1\linewidth} p{0.1\linewidth} p{0.1\linewidth} p{0.1\linewidth} p{0.12\linewidth} p{0.12\linewidth}}
\hline
\multicolumn{9}{c}{\textbf{BEIR Dataset}} \\
\hline
Model & Recall & ArguAna & FIQA & Quora & SciDocs & SciFact & CQADupstack English & CQADupstack Physics \\
\hline

\multirow{2}{*}{ColBERTv2} & recall@3 & 0.3542 & 0.4469 & 0.9048 & 0.2990 & 0.6691 & 0.4484 & 0.4052 \\
& recall@10 & 0.6287 & 0.5787 & 0.9643 & 0.4780 & 0.7755 & 0.5369 & 0.5380 \\
\hline
\multirow{2}{*}{ColBERTv2*} & recall@3 & \textbf{0.3592} & \textbf{0.4666} & \textbf{0.9067} & \textbf{0.3000} & \textbf{0.6862} & \textbf{0.4822} & \textbf{0.4196} \\
& recall@10 & \textbf{0.6344} & \textbf{0.6018} & \textbf{0.9663} & \textbf{0.4850} & \textbf{0.7917} & \textbf{0.5694} & \textbf{0.5611} \\
\hline
\hline
\end{tabular}
}
\caption{Results on ColBERTv2: The performance of LLM-augmented ColBERTv2 has greatly exceeded the performance of vanilla ColBERTv2 on both LoTTE and BEIR dataset. * means base model plus the doc-level embedding. }
\label{tab:colbertv2_results}
\end{table*}

\section{Ablation Studies}
In this section, we want to see how the chunk, query and title fields impact the retrieval quality of different retriever models. For Bi-encoders (Contriever and DRAGON), we further control the field weights of chunk, synthetic query and title to see how those parameters can affect the performance. For the token-level late-interaction model (ColBERTv2), we just control the model to use one of the chunk, query or title field only to see how the they affect the end-to-end retrieval quality. 

For the Contriever model (Table \ref{tab:contriever_ablation}), we have observed that most of the time the synthetic queries play the most critical role in boosting the recall performance, comparing to the other two fields, while in BEIR-SciDocs and BEIR-Scifact, synthetic queries' importance is smaller. Therefore, a weighted sum of multiple fields in doc-level embedding yields better performance in most cases. And these weights can be further tuned as hyper-parameters. 

For the DRAGON model (Table\ref{tab:dragon_ablation}), there’s less strong pattern that which field plays a more important role in doc-level embedding. In the LoTTE dataset it's more driven by the title field. However, the chunk field matters more in datasets of BEIR-ArguAna, BEIR-Quora and BEIR-SciFact. Once again, a weighted sum of multiple document fields in doc-level embedding yields better performance in most cases. For the different patterns we have observed in DRAGON vs Contriever, one reason might be that DRAGON uses separate query and context encoders, while Contriever uses shared query and context encoders in our setup. Thus Contriever is better at identifying similarity instead of relevance and that’s why the synthetic query field has a more significant impact in the Contriever model, since it better transforms similarity to relevance as explained before. 

For ColBERTv2 (Table \ref{tab:colbertv2_ablation}), we have observed from the LoTTE dataset that further chunking the passages actually hurts the performance. This may be because the similarity calculation is done at token-level so chunking the passages (meanwhile increasing the number of chunks) will not help the ColBERTv2 model digest the granular contextual information. Therefore we did not evaluate the chunk-only scenario on the BEIR dataset. The synthetic queries are more critical than titles for ColBERTv2 across the datasets, while combining them all often yields even better recall results. Note again there’s no field weights hyper-parameters for token-level late-interaction models.

\section{Supervised Fine-tuning}
The above studies are all based on zero-shot evaluation, namely the checkpoints in \citep{lin2023train, santhanam2021colbertv2} are directly used in evaluation. Actually our LLM-augmented retrieval and doc-level embedding also support supervised fine-tuning on domain-specific datasets. Some popular training methods include picking hard negatives, constructing in-batch or cross-batch negatives, and calculating InfoNCE loss as training loss \citep{izacard2021unsupervised, lin2023train}. In this section, we propose some effective training techniques for fine-tuning which we used internally on proprietary dataset and performed better than the common methods mentioned above. 

\subsection{Adaptive Negative Sampling}
The process of acquiring negative docs for each query can be approached in several ways. A straightforward method involves the random sampling of doc contexts, excluding the ground truth. Alternatively, a model such as BM25 \citep{robertson2004simple, robertson2009probabilistic} could be employed to select “strong” negative doc contexts that have been assigned high ranking scores by the model. However, for more effective training, it is advantageous to obtain “difficult” negative doc contexts. To this end, an adaptive negative sampling approach was developed. The retrieval model under training is used to rank all doc contexts for each query. The doc contexts ranked in the top positions, excluding the ground truth doc contexts, are selected and then represented as the most challenging negative examples that could potentially confound our retrieval model. Consequently, it is imperative to train the retrieval model to effectively handle these samples. 

A significant challenge of this approach is the time and space complexity. For each query (or batch of queries), it is necessary to re-compute the embedding for all doc contexts to identify the top negative examples. These doc context embedding cannot be pre-calculated due to the continuous updating of the model parameters during training, which correspondingly alters the doc context embedding. To address this issue, we can periodically pre-calculate and update the top negative samples for multiple training batches simultaneously. 

\subsection{Loss Function}
In supervised training, for each training query, each doc has a binary label: relevant or not relevant.

In the context of binary label problems, the cross-entropy loss function is a prevalent choice. However, the cross-entropy is primarily for classification problems. This is due to its requirement for model output scores to be maximized for positive examples and minimized for negative examples. Within the realm of information retrieval, the primary challenge lies in the task of ranking of target documents. The absolute numerical values of output scores are not the primary focus. Instead, the focus is on the relative order and gaps between the model outputs of positive and negative doc contexts. Consequently, we adopted the margin ranking loss function \citep{nayyeri2019adaptive}, which aims to maximize the "margin" between the scores of positive and negative examples.
\begin{equation}
\resizebox{.85\hsize}{!}{$Loss(o_1, o_2, y) = max(0, -y(o_1-o_2) + margin)$}
\end{equation}

In the above equation, $o_1$, $o_2$ are model output scores, the value of $y$ equals 1 or -1, indicating whether $o_1$ should be larger or smaller than $o_2$ respectively. For example, when $y$=1, and the gap between $o_1$ and $o_2$ is larger than the “margin” parameter, the loss value is 0. Otherwise, the smaller the gap between $o_1$ and $o_2$ , the larger the loss value. Our experiments showed that this margin loss was better than the cross entropy loss on proprietary domain data fine-tuning.

\section{Conclusion}
This paper presents a novel framework, LLM-augmented retrieval, which significantly improves the performance of existing retriever models by enriching the embedding of documents through large language model augmentation. The proposed framework includes a doc-level embedding that encodes contextual information from synthetic queries, titles, and chunks, which can be adapted to various retriever model architectures. The proposed approach has achieved state-of-the-art results across different models and datasets, demonstrating its effectiveness in enhancing the quality and robustness of neural information retrieval. Future research could explore further enhancements to the LLM-augmented retrieval framework, such as the integration of additional contextual information into doc-level embedding, the application of more advanced similarity score measures, more complicated approaches to combine the embedding of multiple chunks/queries into one chunk/query field embedding, etc.

\section{Limitations}
One limitation of study include the extra computation resources it required in augmenting relevant queries and titles for original documents and sometimes to size of augmented texts can be comparable to the size of the original documents. This computational limitation may restrict the usage of this approach where computational resource is limited.

Another limitation or risk is that the hallucination in large language models may pose extra inaccuracy in augmented corpus to the original documents. Hallucination remains an unsolved problem in the field of large language model's study.

\bibliography{custom}

\appendix
\section*{Appendix}
\label{sec:appendix}

\begin{table*}[ht!]
\centering
\resizebox{\textwidth}{!}{%
\begin{tabular}{ p{0.15\linewidth} p{0.1\linewidth} p{0.1\linewidth} p{0.1\linewidth} p{0.1\linewidth} p{0.1\linewidth} p{0.1\linewidth} p{0.1\linewidth} p{0.1\linewidth} p{0.1\linewidth} p{0.1\linewidth} p{0.1\linewidth}}
\hline
\multicolumn{12}{c}{\textbf{LoTTE Dataset}} \\
\hline
Model & Recall & Lifestyle Forum & Lifestyle Search & Recreation Forum & Recreation Search & Science Forum & Science Search & Technology Forum & Technology Search & Writing Forum & Writing Search\\
\hline
\multirow{2}{*}{\shortstack {Contriever\\ $w_{chunk}$=1.0}} & recall@3  & 0.4476	 & 0.4342	 & 0.3871	 & 0.2695	 & 0.1358	 & 0.1896	 & 0.2116	 & 0.1879	 & 0.4425	 & 0.3968 \\
& recall@10  & 0.6459	 & 0.6172	 & 0.5415	 & 0.4145	 & 0.2196	 & 0.3063	 & 0.3693	 & 0.3003	 & 0.600	 & 0.5369 \\
\hline
\multirow{2}{*}{\shortstack {Contriever\\ $w_{query}$=1.0}} & recall@3  & 0.6194	 & \textbf{0.6967}	 & 0.5355	 & 0.4437	 & 0.2335	 & 0.2901	 & 0.3523	 & 0.3305	 & 0.5860	 & 0.5472 \\
& recall@10   & 0.7762	 & \textbf{0.7837}	 & 0.6863	 & 0.6115	 & 0.3461	 & 0.4295	 & 0.5180	 & 0.4883	 & 0.7410	 & 0.6910 \\
\hline
\multirow{2}{*}{\shortstack {Contriever\\ $w_{title}$=1.0}} & recall@3   & 0.5310	 & 0.4902	 & 0.4975	 & 0.3789	 & 0.2345	 & 0.1896	 & 0.3468	 & 0.2668	 & 0.5315	 & 0.4809 \\
& recall@10    & 0.6958	 & 0.6641	 & 0.6404	 & 0.5314	 & 0.3421	 & 0.3241	 & 0.5200	 & 0.4077	 & 0.6725	 & 0.6153 \\
\hline
\multirow{2}{*}{Contriever*} & recall@3   & \textbf{0.6244}	 & 0.6021	 & \textbf{0.5455}	 & \textbf{0.4610}	 & \textbf{0.2395}	 & \textbf{0.2901}	 & \textbf{0.3663}	 & \textbf{0.3557}	 & \textbf{0.5970}	 & \textbf{0.5724} \\
& recall@10   & \textbf{0.7622}	 & 0.7821	 & \textbf{0.6948}	 & \textbf{0.6320}	 & \textbf{0.3570}	 & \textbf{0.4684}	 & \textbf{0.5494}	 & \textbf{0.5017}	 & \textbf{0.7365}	 & \textbf{0.6919} \\
\hline
\end{tabular}
}
\resizebox{\textwidth}{!}{%
\begin{tabular}{ p{0.15\linewidth} p{0.1\linewidth} p{0.1\linewidth} p{0.1\linewidth} p{0.1\linewidth} p{0.1\linewidth} p{0.1\linewidth} p{0.12\linewidth} p{0.12\linewidth}}
\hline
\multicolumn{9}{c}{\textbf{BEIR Dataset}} \\
\hline
Model & Recall & ArguAna & FIQA & Quora & SciDocs & SciFact & CQADupstack English & CQADupstack Physics \\
\hline
\multirow{2}{*}{\shortstack {Contriever\\ $w_{chunk}$=1.0}} & recall@3   & 0.2240	 & 0.2623	 & 0.8653	 & 0.1980	 & \textbf{0.6177}	 & 0.2605	 & 0.2156 \\
& recall@10   & 0.5391	 & 0.4031	 & 0.9463	 & 0.3360	 & \textbf{0.7466}	 & 0.3580	 & 0.3292 \\
\hline
\multirow{2}{*}{\shortstack {Contriever\\ $w_{query}$=1.0}} & recall@3   & 0.2347	 & 0.3580	 & 0.8622	 & 0.2180	 & 0.5888	 & \textbf{0.3860}	 & 0.3330 \\
& recall@10    & 0.5718	 & 0.5045	 & 0.8088	 & 0.3720	 & 0.7322	 & 0.5013	 & 0.4629 \\
\hline
\multirow{2}{*}{\shortstack {Contriever\\ $w_{title}$=1.0}} & recall@3    & 0.2063	 & 0.3180	 & 0.7555	 & \textbf{0.2600}	 & 0.5573	 & 0.3338	 & 0.2926 \\
& recall@10     & 0.5192	 & 0.4595	 & 0.8791	 & \textbf{0.4120}	 & 0.7051	 & 0.4369	 & 0.4100 \\
\hline
\multirow{2}{*}{Contriever*} & recall@3    & \textbf{0.2468}	 & \textbf{0.3690}	 & \textbf{0.8687}	 & 0.2440	 & 0.5996	 & 0.3822	 & \textbf{0.3417} \\
& recall@10    & \textbf{0.5825}	 & \textbf{0.5174}	 & \textbf{0.9517}	 & 0.4030	 & 0.7259	 & \textbf{0.5025}	 & \textbf{0.4658} \\
\hline
\end{tabular}
}
\caption{Ablation study on doc-level embedding with Contriever. In most cases the ensemble of relevant queries, title and chunks gives the best results. * means base model plus the doc-level embedding (chunk:0.1, query:1.0, title:0.5).}
\label{tab:contriever_ablation}
\end{table*}

\begin{table*}[ht!]
\centering
\resizebox{\textwidth}{!}{%
\begin{tabular}{ p{0.15\linewidth} p{0.1\linewidth} p{0.1\linewidth} p{0.1\linewidth} p{0.1\linewidth} p{0.1\linewidth} p{0.1\linewidth} p{0.1\linewidth} p{0.1\linewidth} p{0.1\linewidth} p{0.1\linewidth} p{0.1\linewidth}}
\hline
\multicolumn{12}{c}{\textbf{LoTTE Dataset}} \\
\hline
Model & Recall & Lifestyle Forum & Lifestyle Search & Recreation Forum & Recreation Search & Science Forum & Science Search & Technology Forum & Technology Search & Writing Forum & Writing Search\\
\hline
\multirow{2}{*}{\shortstack {DRAGON\\ $w_{chunk}$=1.0}} & recall@3   & 0.6244	 & 0.7126	 & 0.5375	 & 0.5779	 & 0.2385	 & 0.3695	 & 0.3239	 & 0.4362	 & 0.6190	 & 0.6471 \\
& recall@10   & 0.7627	 & 0.8636	 & 0.6813	 & 0.7359	 & 0.3738	 & 0.5462	 & 0.5010	 & 0.6023	 & 0.7620	 & 0.7871 \\
\hline
\multirow{2}{*}{\shortstack {DRAGON\\ $w_{query}$=1.0}} & recall@3   & 0.6583	 & 0.7247	 & 0.5839	 & 0.6071	 & 0.2707	 & 0.3647	 & 0.3892	 & 0.4866	 & 0.6235	 & 0.6583 \\
& recall@10    & 0.8017	 & 0.8654	 & 0.7108	 & 0.7478	 & 0.3991	 & 0.5219	 & 0.5704	 & 0.6812	 & 0.7420	 & 0.7656 \\
\hline
\multirow{2}{*}{\shortstack {DRAGON\\ $w_{title}$=1.0}} & recall@3    & \textbf{0.6913}	 & 0.7610	 & \textbf{0.6294}	 & 0.6472	 & \textbf{0.3565}	 & 0.4408	 & \textbf{0.4616}	 & \textbf{0.5436}	 & \textbf{0.6550}	 & 0.6928 \\
& recall@10    & 0.8167	 & 0.8790	 & 0.7458	 & 0.7879	 & \textbf{0.4834}	 & 0.5948	 & \textbf{0.6477}	 & 0.7064	 & 0.7690	 & 0.8011 \\
\hline
\multirow{2}{*}{DRAGON*} & recall@3    & 0.6883	 & \textbf{0.7625}	 & 0.6079	 & \textbf{0.6472}	 & 0.3099	 & \textbf{0.4498}	 & 0.4192	 & 0.5285	 & 0.6520	 & \textbf{0.7031} \\
& recall@10    & \textbf{0.8172}	 & \textbf{0.8911}	 & \textbf{0.7468}	 & \textbf{0.7944}	 & 0.4427	 & \textbf{0.6062}	 & 0.6038	 & \textbf{0.7097}	 & \textbf{0.7725}	 & \textbf{0.8170} \\
\hline
\end{tabular}
}
\resizebox{\textwidth}{!}{%
\begin{tabular}{ p{0.15\linewidth} p{0.1\linewidth} p{0.1\linewidth} p{0.1\linewidth} p{0.1\linewidth} p{0.1\linewidth} p{0.1\linewidth} p{0.12\linewidth} p{0.12\linewidth}}
\hline
\multicolumn{9}{c}{\textbf{BEIR Dataset}} \\
\hline
Model & Recall & ArguAna & FIQA & Quora & SciDocs & SciFact & CQADupstack English & CQADupstack Physics \\
\hline
\multirow{2}{*}{\shortstack {DRAGON\\ $w_{chunk}$=1.0}} & recall@3    & \textbf{0.3919}	 & 0.3681	 & 0.8587	 & 0.2860	 & 0.6601	 & 0.4331	 & 0.3638 \\
& recall@10    & \textbf{0.6863}	 & 0.5196	 & 0.9478	 & 0.4700	 & \textbf{0.7827}	 & 0.5338	 & 0.4966 \\
\hline
\multirow{2}{*}{\shortstack {DRAGON\\ $w_{query}$=1.0}} & recall@3    & 0.3265	 & 0.3875	 & 0.8267	 & 0.2820	 & 0.6032	 & 0.4318	 & 0.3503 \\
& recall@10     & 0.6472	 & 0.5220	 & 0.9283	 & 0.4470	 & 0.7403	 & 0.5344	 & 0.4889 \\
\hline
\multirow{2}{*}{\shortstack {DRAGON\\ $w_{title}$=1.0}} & recall@3    & 0.3208	 & \textbf{0.4310}	 & 0.8039	 & 0.2940	 & 0.6375	 & 0.4516	 & \textbf{0.4081} \\
& recall@10      & 0.6230	 & \textbf{0.5692}	 & 0.9139	 & 0.4770	 & 0.7556	 & 0.5567	 & 0.5274 \\
\hline
\multirow{2}{*}{DRAGON*} & recall@3     & 0.3663	 & 0.4255	 & \textbf{0.8638}	 & \textbf{0.3040}	 & \textbf{0.6610}	 & \textbf{0.4618}	 & 0.3936 \\
& recall@10     & 0.6764	 & 0.5635	 & \textbf{0.9527}	 & \textbf{0.4800}	 & 0.7710	 & \textbf{0.5662}	 & \textbf{0.5342} \\
\hline
\end{tabular}
}
\caption{Ablation study on doc-level embedding with DRAGON. In most cases the ensemble of relevant queries, title and chunks gives the best results. * means base model plus the doc-level embedding (chunk:0.3, query:0.6, title:0.3).}
\label{tab:dragon_ablation}
\end{table*}

\begin{table*}[ht!]
\centering
\resizebox{\textwidth}{!}{%
\begin{tabular}{ p{0.15\linewidth} p{0.1\linewidth} p{0.1\linewidth} p{0.1\linewidth} p{0.1\linewidth} p{0.1\linewidth} p{0.1\linewidth} p{0.1\linewidth} p{0.1\linewidth} p{0.1\linewidth} p{0.1\linewidth} p{0.1\linewidth}}
\hline
\multicolumn{12}{c}{\textbf{LoTTE Dataset}} \\
\hline
Model & Recall & Lifestyle Forum & Lifestyle Search & Recreation Forum & Recreation Search & Science Forum & Science Search & Technology Forum & Technology Search & Writing Forum & Writing Search\\
\hline
\multirow{2}{*}{\shortstack {ColBERTv2\\ chunk64 only}} & recall@3   & 0.6184	 & 0.7474	 & 0.5949	 & 0.6245	 & 0.3535	 & 0.4797	 & 0.3927	 & 0.5168	 & 0.6680	 & 0.7199 \\
& recall@10  & 0.7537	 & 0.8759	 & 0.7258	 & 0.7586	 & 0.4819	 & 0.6353	 & 0.5758	 & 0.6846	 & 0.7945	 & 0.8273 \\
\hline
\multirow{2}{*}{\shortstack {ColBERTv2\\ query only}} & recall@3   & 0.7088	 & 0.7413	 & 0.6479	 & 0.6580	 & 0.3634	 & 0.4327	 & 0.3643	 & 0.4530	 & 0.6835	 & 0.7274 \\
& recall@10    & 0.8222	 & 0.8759	 & 0.7642	 & 0.7727	 & 0.4948	 & 0.5997	 & 0.5259	 & 0.5419	 & 0.7890	 & 0.8254 \\
\hline
\multirow{2}{*}{\shortstack {ColBERTv2\\ title only}} & recall@3    & 0.6004	 & 0.6218	 & 0.5210	 & 0.5487	 & 0.3128	 & 0.3695	 & 0.4336	 & 0.4715	 & 0.5425	 & 0.5780 \\
& recall@10     & 0.7368	 & 0.7458	 & 0.6479	 & 0.6937	 & 0.4378	 & 0.5024	 & 0.5968	 & 0.6141	 & 0.6505	 & 0.6853 \\
\hline
\multirow{2}{*}{ColBERTv2*} & recall@3    & \textbf{0.7308}	 & \textbf{0.8003}	 & \textbf{0.6753}	 & \textbf{0.7100 }& \textbf{0.4026}	 & \textbf{0.5024}	 & \textbf{0.4626}	 & \textbf{0.5956}	 & \textbf{0.7145}	 & \textbf{0.7544} \\
& recall@10    & \textbf{0.8447}	 & \textbf{0.9107}	 & \textbf{0.7862}	 & \textbf{0.8268}	 & \textbf{0.5558}	 & \textbf{0.6726}	 & \textbf{0.6517}	 & \textbf{0.7383}	 & \textbf{0.8260} & \textbf{0.8571} \\
\hline
\end{tabular}
}
\resizebox{\textwidth}{!}{%
\begin{tabular}{ p{0.15\linewidth} p{0.1\linewidth} p{0.1\linewidth} p{0.1\linewidth} p{0.1\linewidth} p{0.1\linewidth} p{0.1\linewidth} p{0.12\linewidth} p{0.12\linewidth}}
\hline
\multicolumn{9}{c}{\textbf{BEIR Dataset}} \\
\hline
Model & Recall & ArguAna & FIQA & Quora & SciDocs & SciFact & CQADupstack English & CQADupstack Physics \\
\hline
\multirow{2}{*}{\shortstack {ColBERTv2\\ query only}} & recall@3    & 0.3122	 & 0.4299	 & 0.8037	 & 0.2680	 & 0.6041	 & 0.4503	 & 0.4187 \\
& recall@10     & 0.5711	 & 0.5654	 & 0.9102	 & 0.4170	 & 0.7214	 & 0.5357	 & 0.5342 \\
\hline
\multirow{2}{*}{\shortstack {ColBERTv2\\ title only}} & recall@3     & 0.2091	 & 0.3372	 & 0.7149	 & 0.2580	 & 0.4806	 & 0.3344	 & 0.3494 \\
& recall@10     & 0.3947	 & 0.4588	 & 0.8265	 & 0.4060	 & 0.6005	 & 0.4248	 & 0.4716 \\
\hline
\multirow{2}{*}{ColBERTv2*} & recall@3     & \textbf{0.3592}	 & \textbf{0.4666}	 & \textbf{0.9067}	 & \textbf{0.3000} & \textbf{0.6862}	 & \textbf{0.4822}	 & \textbf{0.4196} \\
& recall@10     & \textbf{0.6344}	 & \textbf{0.6018}	 & \textbf{0.9663}	 & \textbf{0.4850}	 & \textbf{0.7917}	 & \textbf{0.5694}	 & \textbf{0.5611} \\
\hline
\end{tabular}
}
\caption{Ablation study on doc-level embedding with ColBERTv2. In all cases the ensemble of relevant queries, title and chunks gives the best results. * means base model plus the doc-level embedding.}
\label{tab:colbertv2_ablation}
\end{table*}

\begin{table*}
\centering
\resizebox{\columnwidth}{!}{%
\begin{tabular}{l l l l }
\hline
\textbf{Question Set} & \textbf{\#Questions} & \textbf{\#Passages} \\
\hline
ArguAna & 1406 & 8674 \\
\hline
FIQA & 6648 & 57600 \\
\hline
Quora & 15000 & 522929 \\
\hline
SciDocs & 1000 & 25313 \\
\hline
SciFact & 1109 & 5183 \\
\hline
CQADupstack English & 1570 & 40221 \\
\hline
CQADupstack Physics & 1039 & 38316 \\
\hline
\end{tabular}
\begin{tabular}{lc}
\hline
\end{tabular}
}
\caption{Statistical description of BEIR dataset}
\label{tab:beir}
\end{table*}

\begin{table*}
\centering
\resizebox{\columnwidth}{!}{%
\begin{tabular}{l l l l }
\hline
\textbf{Question Set} & \textbf{\#Questions} & \textbf{\#Passages} & \textbf{Subtopics}\\
\hline
Lifestyle Search & 661 & \multirow{2}{*}{219k} & \multirow{2}{*}{Cooking, Sports, Travel} \\
Lifestyle Forum & 2002 \\
\hline
Recreation Search & 924 & \multirow{2}{*}{167k} & \multirow{2}{*}{Gaming, Anime, Movies} \\
Recreation Forum & 2002 \\
\hline
Science Search & 617 & \multirow{2}{*}{1.694M} & \multirow{2}{*}{Math, Physics, Biology} \\
Science Forum & 2017 \\
\hline
Technology Search & 596 & \multirow{2}{*}{639k} & \multirow{2}{*}{Apple, Android, UNIX, Security} \\
Technology Forum & 2004 \\
\hline
Writing Search & 1071 & \multirow{2}{*}{200k} & \multirow{2}{*}{English} \\
Writing Forum & 2000 \\
\hline
\end{tabular}
\begin{tabular}{lc}
\hline
\end{tabular}
}
\caption{Statistical description of LoTTE Test dataset}
\label{tab:lotte}
\end{table*}

Here we share the ablation studies' results and the prompts we used in generating synthetic queries and titles using Llama-70B.
\begin{table*}[h!]
\begin{tabular}{p{1\linewidth}}
\hline
I will give you an article below. What are some search queries or questions that are relevant for this article or this article can answer? \\
Separate each query in a new line. \\
This is the article: \{document\} \\
Only provide the user queries without any additional text. Format every query as 'query:' followed by the question. Don't write empty queries. \\
\hline
\end{tabular}
\caption{Prompt for generating relevant queries for documents}
\label{tab:prompt_query}
\end{table*}

\begin{table*}[h!]
\begin{tabular}{p{1\linewidth}}
\hline
I will give you an article below. Create a title for the below article. \\
This is the article: \{document\} \\
Only provide the title without any additional text. Format the reply starting with 'title:' followed by the question. Don't write empty title. \\
\hline
\end{tabular}
\caption{Prompt for generating titles for documents.}
\label{tab:prompt_title}
\end{table*}

\end{document}